\documentstyle[12pt,epsfig]{article}
\begin{document}

\begin{flushright}
\underline{BIHEP--TH--2002--51} \\
November 2002
\end{flushright}

\vspace{0.2cm}

\begin{center}
{\Large\bf The LOW Solution and Two-zero Textures \\
of the Neutrino Mass Matrix}
\end{center}

\vspace{0.2cm}

\begin{center}
{\bf Wan-lei Guo} 
\footnote{Electronic address: guowl@mail.ihep.ac.cn} 
~ and ~
{\bf Zhi-zhong Xing} 
\footnote{Electronic address: xingzz@mail.ihep.ac.cn}
\end{center}
\begin{center}
{\it Institute of High Energy Physics, Chinese Academy of Sciences, \\
P.O. Box 918 (4), Beijing 100039, China}
\end{center}

\vspace{2cm}

\begin{abstract}
As a viable alternative to the LMA solution to the solar neutrino
problem, the LOW solution is confronted with two-zero textures of 
the $3\times 3$ neutrino mass matrix. We find out nine acceptable 
textures, from which instructive predictions can be obtained for the 
absolute values of neutrino masses, Majorana phases of CP violation, 
and effective masses of the tritium beta decay and neutrinoless 
double beta decay. 
\end{abstract}

\newpage

\framebox{\Large\bf 1} ~
In the standard electroweak model, three known neutrinos are assumed to 
be the exactly massless Weyl particles. This assumption has no conflict 
with all of today's direct-mass-search experiments \cite{PDG},
but it is not guaranteed by any fundamental symmetry principle of particle 
physics. Indeed most extensions of the standard model, such as the SO(10)
grand unified theories \cite{Altarelli}, do allow neutrinos to have tiny 
masses. If three neutrinos have non-degenerate masses, their mass eigenstates 
$(\nu_1, \nu_2, \nu_3)$ may not coincide with their flavor eigenstates 
$(\nu_e, \nu_\mu, \nu_\tau)$, leading to lepton flavor mixing.

The recent SNO results \cite{SNO} provide compelling evidence that solar 
neutrinos undergo the flavor conversion 
($\nu_e \rightarrow \nu_\mu, \nu_\tau$) during their travel to the earth. 
This anomaly, similar to the observed deficit of atmospheric $\nu_\mu$
neutrinos \cite{ATM}, is most likely due to neutrino oscillations --
a quantum phenomenon which can naturally happen if neutrinos are massive
and lepton flavors are mixed. In the framework of neutrino oscillations, 
a global analysis of current solar neutrino data indicates that the
LMA solution is most favored \cite{Smirnov}. As a viable alternative to
the LMA solution, the LOW solution is accepted at about $3\sigma$ 
level \cite{LOW}. Before the LMA solution is firmly established as the
correct solution to the solar neutrino problem, it is certainly
meaningful and useful to study the LOW solution (as well as other 
alternatives) and its implication on neutrino masses and lepton flavor 
mixing parameters.

In this paper, we aim to confront the LOW solution with two-zero textures
of the $3\times 3$ neutrino mass matrix $M$ in the flavor basis where the
charged lepton mass matrix is diagonal. The reason why we take 
into account texture zeros of $M$ is simple: $M$ totally involves nine
physical parameters (three neutrino masses, three flavor mixing angles,
and three CP-violating phases), but only six of them or their combinations
(two neutrino mass-squared differences, three mixing angles and one
CP-violating phase) can in principle be determined from neutrino 
oscillations. To recast the structure of $M$ from current experimental
data, two or more extra constraints have to be assumed, either for 
some elements of $M$ (approach A) or for neutrino masses and CP-violating
phases (approach B). Some recent works on texture zeros of $M$ done by
a number of authors \cite{Glashow,Xing02,Tanimoto,Desai} belong to approach A, 
while the systematic analysis of $M$ shown in Ref. \cite{Smirnov2} belongs
to approach B. Such phenomenological attempts 
are important, because they may shed light on the underlying flavor
symmetry and its breaking mechanism responsible for the structure of $M$,
from which it is likely to obtain a deeper insight into the generation of 
neutrino masses and lepton flavor mixing. 

The present work is different from the previous 
ones \cite{Glashow,Xing02,Tanimoto,Desai}
in several aspects. First of all, we concentrate on the LOW solution
instead of the LMA solution. The former has not been confronted with
two-zero textures of the neutrino mass matrix $M$ in the literature. Second, 
we carry out a careful numerical analysis of every two-zero pattern of $M$ 
to pin down its complete parameter space, because simple analytical 
approximations are sometimes unable to reveal the whole regions of relevant
parameters allowed by current experimental data. Third, we quantitatively
obtain the ranges of the absolute neutrino masses, the Majorana phases of 
CP violation, and the effective masses of the tritiun beta decay and 
neutrinoless double beta decay. We find that nine of the fifteen two-zero
textures of $M$ are compatible with the LOW solution, although two of them
are only marginally allowed. 

\vspace{0.3cm}

\framebox{\Large\bf 2} ~
In the flavor basis where the charged lepton mass matrix is
diagonal, the Majorana neutrino mass matrix can be expressed as
\begin{equation}
M \; =\; U \left ( \matrix{
\lambda_1 & 0 & 0 \cr
0 & \lambda_2 & 0 \cr
0 & 0 & \lambda_3 \cr} \right ) U^{\rm T} \; ,
\end{equation}
where $\lambda_i$ (for $i=1,2,3$) stand for the neutrino mass eigenvalues
consisting of two nontrivial CP-violating phases ($\rho$ and 
$\sigma$), and $U$ is a CKM-like flavor mixing matrix containing
three rotation angles ($\theta_x$, $\theta_y$ and $\theta_z$) and
another CP-violating phase ($\delta$) \cite{FX98}. Without loss of generality, 
we take the phase convention
\begin{equation}
\lambda_1 = m_1 e^{2i\rho} \; , ~~~
\lambda_2 = m_2 e^{2i\sigma} \; , 
\end{equation}
and $\lambda_3 = m_3$ with $m_i$ being the physical neutrino masses,
and parametrize $U$ as
\begin{equation}
U \; = \; \left ( \matrix{
c_x c_z & s_x c_z & s_z \cr
- c_x s_y s_z - s_x c_y e^{-i\delta} &
- s_x s_y s_z + c_x c_y e^{-i\delta} &
s_y c_z \cr 
- c_x c_y s_z + s_x s_y e^{-i\delta} & 
- s_x c_y s_z - c_x s_y e^{-i\delta} & 
c_y c_z \cr } \right ) \; ,
\end{equation}
where $s_x \equiv \sin\theta_x$, $c_x \equiv \cos\theta_x$, and so on.
In this parametrization, the neutrinoless double beta decay is 
associated with $\rho$ and $\sigma$, while CP violation in neutrino 
oscillations depends on $\delta$. Note that three mixing angles
$(\theta_x, \theta_y, \theta_z)$ can all be arranged to lie in the
first quadrant. Arbitrary values between $0$ and $2\pi$ (or
between $-\pi$ and $+\pi$) are allowed for three CP-violating phases 
$(\delta, \rho, \sigma)$.

The symmetric neutrino mass matrix $M$ totally has six independent 
complex entries. If two of them vanish, i.e.,
$M_{ab} = M_{pq} =0$, we obtain two constraint equations:
\begin{eqnarray}
M_{ab} & = & \sum_{i=1}^{3} \left (U_{ai} U_{bi} \lambda_i \right ) = 0 \; , 
\nonumber \\
M_{pq} & = & \sum_{i=1}^{3} \left (U_{pi} U_{qi} 
\lambda_i \right ) = 0 \; ,
\end{eqnarray}
where $a$, $b$, $p$ and $q$ run over $e$, $\mu$ and $\tau$, but $(p,q) \neq (a,b)$.
Solving Eq. (4), we arrive at \cite{Xing02}
\begin{eqnarray}
\frac{\lambda_1}{\lambda_3} & = &
\frac{U_{a3} U_{b3} U_{p2} U_{q2} - U_{a2} U_{b2} U_{p3}
U_{q3}}{U_{a2} U_{b2} U_{p1} U_{q1} - U_{a1} U_{b1}U_{p2} U_{q2}} \; ,
\nonumber \\
\frac{\lambda_2}{\lambda_3} & = &
\frac{U_{a1} U_{b1} U_{p3} U_{q3} - U_{a3} U_{b3} U_{p1}
U_{q1}}{U_{a2} U_{b2} U_{p1} U_{q1} - U_{a1} U_{b1}U_{p2} U_{q2}} \; .
\end{eqnarray}
This result implies that two neutrino mass ratios
$(m_1/m_3, m_2/m_3)$ and two Majorana-type CP-violating phases 
$(\rho, \sigma)$ can fully be determined in terms of three mixing angles 
$(\theta_x, \theta_y, \theta_z)$ and the Dirac-type CP-violating phase 
$(\delta)$. Thus one may examine whether a two-zero texture of $M$ is 
empirically acceptable or not by comparing its prediction for the
ratio of two neutrino mass-squared differences with the result extracted
from current experimental data on solar and atmospheric neutrino 
oscillations:
\begin{equation}
R_\nu \; \equiv \; \frac{\left |m^2_2 - m^2_1 \right |}
{\left |m^2_3 - m^2_2 \right |} \; \approx \; 
\frac{\Delta m^2_{\rm sun}}{\Delta m^2_{\rm atm}} \; .
\end{equation}
The size of $R_\nu$ depends on which solution to the solar neutrino 
problem is taken. 

As for the LOW solution, we have 
$3.5 \times 10^{-8} ~ {\rm eV^2} < \Delta m^2_{\rm sun} <
1.2 \times 10^{-7} ~ {\rm eV^2}$ versus
$1.2 \times 10^{-3} ~ {\rm eV^2} \leq \Delta m^2_{\rm atm}
\leq 4.8 \times 10^{-3} ~ {\rm eV^2}$ 
at the $3\sigma$ confidence level \cite{LOW}, which leads to
$7.3 \times 10^{-6} < R_\nu < 1.0 \times 10^{-4}$. The mixing angles 
of solar and atmospheric neutrino oscillations read as 
$0.43 \leq \tan^2\theta_{\rm sun} \leq 0.86$ and 
$0.3 \leq \sin^2\theta_{\rm atm} \leq 0.7$, obtained from the
same global analysis \cite{LOW}. We then arrive at the ranges
of $\theta_x$ ($\approx \theta_{\rm sun}$) and $\theta_y$
($\approx \theta_{\rm atm}$):
$33.2^\circ < \theta_x < 42.8^\circ$ and 
$33.2^\circ \leq \theta_y \leq 56.8^\circ$. The third mixing angle
$\theta_z$ is restricted by the CHOOZ experiment \cite{CHOOZ}:
$\theta_z \approx \theta_{\rm chz} < 13.3^\circ$ extracted from
the upper limit $\sin^2 2\theta_{\rm chz} < 0.2$. There is no
experimental constraint on the CP-violating phase $\delta$. Hence
we take $\delta$ from $0^\circ$ to $360^\circ$ in our numerical
calculations.

\vspace{0.3cm}

\framebox{\Large\bf 3} ~
There are totally fifteen distinct topologies for the structure of $M$
with two independent vanishing entries, as shown in Tables 1 and 2.
We work out the explicit expressions of $\lambda_1/\lambda_3$ and 
$\lambda_2/\lambda_3$ for each pattern of $M$ by use of Eq. (5), and 
list the results in the same tables. With the input values of 
$\theta_x$, $\theta_y$, $\theta_z$ and $\delta$ mentioned above, we 
calculate the ratio $R_\nu$ and examine whether it is in the range
allowed by current data. This criterion has been used 
before \cite{Glashow,Xing02} to pick the phenomenologically favored
patterns of $M$ in the LMA case. 

Nine of the fifteen two-zero textures of $M$ listed in Table 1
are found to be in accord with the LOW solution as well as the 
atmospheric neutrino data. They can be classified into four categories
\footnote{Note that categories A, B and C correspond to those given 
in Refs. \cite{Glashow,Xing02} for the LMA solution.}:
A (with $\rm A_1$ and $\rm A_2$), B (with $\rm B_1$, $\rm B_2$, $\rm B_3$ 
and $\rm B_4$), C, and D (with $\rm D_1$ and $\rm D_2$). The point of
this classification is that the textures of $M$ in each category result
in similar physical consequences, which are almost indistinguishable 
in practice. The other six patterns of $M$ (categories E and F) 
listed in Table 2 cannot coincide with current experimental data. 
In particular, the exact neutrino mass degeneracy ($m_1 = m_2 = m_3$) 
is predicted from three textures of $M$ belonging to category F. 

Now let us focus on the nine phenomenologically acceptable textures 
of $M$. As $R_\nu$ is required to be very small, the space of four
input parameters ($\theta_x$, $\theta_y$, $\theta_z$ and $\delta$)
may strongly be constrained for a specific two-zero pattern of $M$.
To be more concrete, we take patterns $\rm A_1$, $\rm B_1$, $\rm C$ and 
$\rm D_1$ as four typical examples for numerical illustration. Our
results for $\sin^2 2\theta_{\rm chz}$ versus $\delta$ and $\theta_y$
versus $\theta_x$ are shown Figs. 1 -- 4. Some comments are in order.

(1) For pattern $\rm A_1$, arbitrary values of $\delta$ are allowed
if $\sin^2 2\theta_{\rm chz}$ is tiny ($\leq 0.002$). When the
values of $\delta$ are taken to be around $180^\circ$, however,
$\sin^2 2\theta_{\rm chz}$ can be large enough, up to its 
experimental upper limit. The mixing angles $\theta_x$ and $\theta_y$ 
may take any values in the ranges allowed by current data. Therefore
we conclude that pattern $\rm A_1$ is favored in phenomenology
with little fine-tuning. A similar conclusion can be drawn for
pattern $\rm A_2$.

(2) For pattern $\rm B_1$, $\delta$ is essentially unconstrained 
if $\sin^2 2\theta_{\rm chz}$ is extremely close to zero; and only 
$\delta \approx 90^\circ$ or $\delta \approx 270^\circ$ is acceptable 
if $\sin^2 2\theta_{\rm chz}$ deviates somehow from zero. Except
$\theta_y \neq 45^\circ$, there is no further constraint on the
parameter space of $(\theta_x, \theta_y)$. We conclude that
pattern $\rm B_1$ with maximal CP violation (i.e., 
$\sin\delta \approx \pm 1$) is phenomenologically favored. So
are patterns $\rm B_2$, $\rm B_3$ and $\rm B_4$.

(3) For pattern $\rm C$, $\delta = 90^\circ$ or
$\delta = 270^\circ$ is forbidden. Furthermore, 
$\theta_y = 45^\circ$ is forbidden. We see that the allowed
parameter space of $(\delta, \theta_{\rm chz})$ and that of
$(\theta_x, \theta_y)$ are rather large. Hence pattern C is
also favored in phenomenology.

(4) For pattern $\rm D_1$, $\delta$ is restricted to be around
$0^\circ$ or $360^\circ$. In particular, the region 
$90^\circ \leq \delta \leq 270^\circ$ is entirely excluded.
$\sin^2 2\theta_{\rm chz} > 0.012$ holds for the allowed range
of $\delta$. Different from patterns $\rm A_1$, $\rm B_1$ and
$\rm C$, pattern $\rm D_1$ requires relatively strong correlation 
between $\theta_x$ and $\theta_y$ (e.g., small values of $\theta_y$
are associated with large values of $\theta_x$ in the allowed
parameter space). In this sense, we argue that pattern $\rm D_1$
is less natural in phenomenology, although it has not been
ruled out by current experimental data. A similar argument
can be made for pattern $\rm D_2$. 

At this point, it is worthwhile to compare between LOW and LMA 
solutions against two-zero patterns of the neutrino mass matrix 
$M$. The main difference between two solutions is in their values 
of $R_\nu$; i.e., $R_\nu \sim {\cal O}(10^{-2})$ for LMA and 
$R_\nu \sim {\cal O}(10^{-5})$ for LOW. Hence a specific 
two-zero texture of $M$ may be in accord with both LMA and LOW
solutions, although the relevant parameter space for LOW is
usually smaller than that for LMA (in particular, when the input 
parameters $\theta_{\rm sun}$, $\theta_{\rm atm}$ and 
$\theta_{\rm chz}$ take values at the same confidence level for
two solutions). A careful numerical analysis of the LMA solution
shows that patterns $\rm D_1$ and $\rm D_2$ can marginally be
allowed \cite{Guo}, like the LOW case. It is difficult to observe
this point from simple analytical approximations made in 
Refs. \cite{Glashow,Xing02}. In the spirit of naturalness, however,
we expect that categories $\rm A$, $\rm B$ and $\rm C$ of two-zero 
patterns of $M$ are more favorable than category $\rm D$ in either
LMA or LOW case.

\vspace{0.3cm}

\framebox{\Large\bf 4} ~
A two-zero texture of $M$ has a number of interesting predictions,
in particular, for the absolute neutrino masses and the Majorana
phases of CP violation \cite{Xing02}. With the help of Eq. (5),
one may calculate the mass ratios $m_1/m_3 = |\lambda_1/\lambda_3|$
and $m_2/m_3 = |\lambda_2/\lambda_3|$ as well as the Majorana phases
$\rho = \arg (\lambda_1/\lambda_3)/2$ and
$\sigma = \arg (\lambda_2/\lambda_3)/2$. The absolute neutrino mass
$m_3$ can be determined from 
\begin{equation}
m_3 \; =\; \frac{1}{\sqrt{\displaystyle \left | 1 - 
\left (\frac{m_2}{m_3} \right )^2 \right |}} 
~ \sqrt{\Delta m^2_{\rm atm}} \;\; .
\end{equation}
Therefore a full determination of the mass spectrum of three neutrinos
is actually possible. Then we may obtain definite predictions for the
effective mass of the tritium beta decay,
\begin{equation}
\langle m\rangle_e \; =\; m_1 c^2_x c^2_z + m_2 s^2_x c^2_z +
m_3 s^2_z \; ;
\end{equation}
and that of the neutrinoless double beta decay,
\begin{equation}
\langle m\rangle_{ee} \; =\; \left | m_1 c^2_x c^2_z e^{2i\rho}
+ m_2 s^2_x c^2_z e^{2i\sigma} + m_3 s^2_z \right | \; .
\end{equation}
It is clear that the Dirac phase $\delta$ has no contribution to 
$\langle m\rangle_{ee}$. Note that CP- and T-violating asymmetries
in normal neutrino oscillations are controlled by $\delta$ or the 
rephasing-invariant parameter 
$J = s_x c_x s_y c_y s_z c^2_z \sin\delta$ \cite{FX00}. Because
of the smallness of $\Delta m^2_{\rm sun}$ in the LOW case,
however, there is no hope to measure leptonic CP violation in 
the terrestrial long-baseline neutrino oscillation 
experiments \cite{Barenboim}. Whether $\langle m\rangle_e$ and 
$\langle m\rangle_{ee}$ can be measured remains an open question.
The present experimental upper bounds are 
$\langle m\rangle_e < 2.2 ~ {\rm eV}$ \cite{PDG} and
$\langle m\rangle_{ee} < 0.35 ~ {\rm eV}$ \cite{HM} at the $90\%$
confidence level. The proposed KATRIN experiment \cite{KATRIN}  
is possible to reach the sensitivity 
$\langle m\rangle_e \sim 0.3 ~ {\rm eV}$, and a number of 
next-generation experiments for the neutrinoless double beta
decay \cite{DB} is possible to probe $\langle m\rangle_{ee}$ at
the level of 10 meV to 50 meV.

We perform a numerical calculation of $m_2/m_3$ versus $m_1/m_3$, 
$\sigma$ versus $\rho$, $\langle m\rangle_{ee}$ versus 
$\langle m\rangle_e$, and $J$ versus $m_3$ for patterns $\rm A_1$,
$\rm B_1$, $\rm C$ and $\rm D_1$. The results are shown in 
Figs. 1 -- 4. Some discussions are in order.

(1) For pattern $\rm A_1$, $\rho \approx \delta/2$ or
$\rho \approx \delta/2 - 180^\circ$ and 
$\sigma \approx \rho \pm 90^\circ$ hold. Two neutrino mass ratios 
lie in the ranges $0.001 \leq m_1/m_3 \leq 0.3$ and 
$0.003 \leq m_2/m_3 \leq 0.3$, and the absolute value of $m_3$ is 
in the range $0.035 ~ {\rm eV} \leq m_3 \leq 0.071 ~ {\rm eV}$. As 
$\langle m\rangle_{ee} = 0$ is a direct consequence of texture 
$\rm A_1$, we calculate the sum of three neutrino masses
$\sum m_i$ instead of $\langle m\rangle_{ee}$. The result is
$0.035 ~ {\rm eV} \leq \sum m_i \leq 0.11 ~ {\rm eV}$, in contrast
with $7.13 \times 10^{-5} ~ {\rm eV} \leq \langle m\rangle_e \leq
0.022 ~ {\rm eV}$. The rephasing invariant of CP violation $J$ is 
found to lie in the range $-0.049 \leq J \leq 0.048$. Similar 
predictions are expected for pattern $\rm A_2$.

(2) For pattern $\rm B_1$, $\rho \approx \sigma \approx \delta
- 90^\circ$ or $\rho \approx \sigma \approx \delta - 270^\circ$ 
holds in most cases; and $\sigma \approx -\rho$
holds when $\delta$ is restricted to equal $90^\circ$ or $270^\circ$.
Two neutrino mass ratios may lie either in the range
$0.42 \leq m_1/m_3 \approx m_2/m_3 \leq 0.97$ or in the range
$1.03 \leq m_1/m_3 \approx m_2/m_3 \leq 2.69$, and the value of $m_3$ 
is found to be in the range 
$0.017 ~ {\rm eV} \leq m_3 \leq 0.28 ~ {\rm eV}$. Furthermore, we 
arrive at $0.017 ~ {\rm eV} \leq \langle m\rangle_e 
\approx \langle m\rangle_{ee} \leq 0.27 ~ {\rm eV}$ as well as
$-0.049 \leq J \leq 0.053$. Similar results can be obtained for
patterns $\rm B_2$, $\rm B_3$ and $\rm B_4$.

(3) For pattern $\rm C$, $\sigma \approx \rho$ when $\theta_y$
approaches $45^\circ$; and there is no clear correlation between
$\rho$ and $\sigma$ for other values of $\theta_y$. Two neutrino
mass ratios may be either in the range
$0.92 \leq m_1/m_3 \approx m_2/m_3 \leq 0.99$ or in the range
$1.01 \leq m_1/m_3 \approx m_2/m_3 \leq 11.68$, and the value of $m_3$ 
is found to lie in the range 
$0.003 ~ {\rm eV} \leq m_3 \leq 0.324 ~ {\rm eV}$. It is remarkable
that $\langle m\rangle_{ee} \approx m_3$ holds to a good degree of
accuracy in the allowed space of those input parameters. We also
obtain $0.035 ~ {\rm eV} \leq \langle m\rangle_e \leq 0.330 ~ {\rm eV}$
and $-0.052 \leq J \leq 0.053$.

(4) For pattern $\rm D_1$, $\rho \approx \delta - 90^\circ$ or
 $\rho \approx \delta - 270^\circ$ and
$\sigma \approx \rho \pm 90^\circ$ hold. Two neutrino mass ratios 
lie in the range $2.77 \leq m_1/m_3 \approx m_2/m_3 \leq 27.75$,
and the absolute value of $m_3$ is in the range 
$0.002 ~ {\rm eV} \leq m_3 \leq 0.026 ~ {\rm eV}$. As for the
tritium beta decay and neutrinoless double beta decay, we obtain
$0.034 ~ {\rm eV} \leq \langle m\rangle_e \leq 0.071 ~ {\rm eV}$ and
$0.003 ~ {\rm eV} \leq \langle m\rangle_{ee} \leq 0.020 ~ {\rm eV}$.
The range of $J$ is found to be $-0.049 \leq J \leq 0.050$. Similar
predictions can straightforwardly be made for pattern $\rm D_2$.

We see that there is no hope to measure both $\langle m\rangle_e$
and $\langle m\rangle_{ee}$, if the neutrino mass matrix $M$ takes 
pattern $\rm A_1$ or $\rm A_2$. It is also impossible to detect
$\langle m\rangle_e$ (and extremely difficult to observe 
$\langle m\rangle_{ee}$), if $M$ takes pattern $\rm D_1$ or $\rm D_2$.
As for categories B and C of $M$, the upper limit of 
$\langle m\rangle_e$ can be close to the sensitivity of the
KATRIN experiment ($\sim 0.3 ~ {\rm eV}$ \cite{KATRIN}), 
and that of $\langle m\rangle_{ee}$ is just below the current 
experimental bound \cite{HM}. Although the magnitude of the 
CP-violating parameter $J$ can be as large as $5\%$ in the LOW case, 
it is hopeless to measure leptonic CP violation in any terrestrial 
experiments of neutrino oscillations. 

\vspace{0.3cm}

\framebox{\Large\bf 5} ~
In summary, we have confronted the LOW solution with two-zero textures
of the neutrino mass matrix $M$ in the flavor basis where the charged
lepton mass matrix is diagonal. Nine patterns of $M$, which can be
classified into four distinct categories, are found to be acceptable
in phenomenology. Compared with categories A, B and C, category D seems
less favored by current experimental data. This situation is similar 
to the LMA case. We expect that new data to be accumulated from solar 
and atmospheric neutrino experiments will allow us to isolate fewer 
two-zero patterns of $M$ which are phenomenologically favored. 

Of course, to pin down the correct solution to the solar neutrino
problem is urgent and important. We have shown that LOW and LMA
solutions are essentially compatible with the same textures of $M$,
although the parameter space for the former is usually smaller
or a bit contrived. One may carry out a similar analysis for the
VO solution to the solar neutrino problem, which has  
$R_\nu \sim 10^{-7}$ \cite{Smirnov}, by use of the analytical results 
presented in Tables 1 and 2. In this case, some fine-tuning of the input 
parameters is unavoidable to make $R_\nu$ strongly suppressed. Once we 
are aware of the true solution, some better understanding of two-zero 
textures of the neutrino mass matrix will be available.

Finally it is worth remarking that a specific texture of lepton mass 
matrices may not be preserved to all orders or at any energy scales in 
the unspecified interactions from which lepton masses are generated. 
Nevertheless, those phenomenologically favored textures at low energy 
scales, no matter whether they are of the two-zero form or other 
forms \cite{Others1,Others2}, are possible to provide enlightening hints 
at the underlying dynamics of lepton mass generation at high energy scales. 


\newpage

\begin{table}
\caption{Nine patterns of the neutrino mass matrix $M$ with two independent
vanishing entries, which are {\it compatible} with the LOW solution and other
empirical hypotheses. The analytical results for two ratios of three neutrino 
mass eigenvalues $\lambda_1/\lambda_3$ and $\lambda_2/\lambda_3$ are 
given in terms of four flavor mixing parameters $\theta_x$, $\theta_y$, 
$\theta_z$ and $\delta$.}
\begin{center}
\begin{tabular}{lcl} \hline\hline 
Pattern of $M$ &~~& Results of $\lambda_1/\lambda_3$ and 
$\lambda_2/\lambda_3$ \\ \hline \\
$\rm A_1: 
\left ( \matrix{
{\bf 0} & {\bf 0} & \times \cr
{\bf 0} & \times & \times \cr
\times & \times & \times \cr} \right )$
&& 
{\large
$\matrix{
\frac{\lambda_1}{\lambda_3} =  
+ \frac{s_z}{c^2_z} \left ( \frac{s_x s_y}{c_x c_y} ~ e^{i\delta} 
- s_z \right ) \cr
\frac{\lambda_2}{\lambda_3} =  
- \frac{s_z}{c^2_z} \left ( \frac{c_x s_y}{s_x c_y} ~ e^{i\delta} 
+ s_z \right ) }$
}
\\ \\
$\rm A_2: 
\left ( \matrix{
{\bf 0} & \times & {\bf 0} \cr
\times & \times & \times \cr
{\bf 0} & \times & \times \cr} \right )$
&& 
{\large
$\matrix{
\frac{\lambda_1}{\lambda_3} =  
- \frac{s_z}{c^2_z} \left ( \frac{s_x c_y}{c_x s_y} ~ e^{i\delta} 
+ s_z \right ) \cr
\frac{\lambda_2}{\lambda_3} =  
+ \frac{s_z}{c^2_z} \left ( \frac{c_x c_y}{s_x s_y} ~ e^{i\delta} 
- s_z \right ) }$
}
\\ \\
$\rm B_1: 
\left ( \matrix{
\times & \times & {\bf 0} \cr
\times & {\bf 0} & \times \cr
{\bf 0} & \times & \times \cr} \right )$
&& 
{\large
$\matrix{
\frac{\lambda_1}{\lambda_3} =  
\frac{s_x c_x s_y \left (2 c^2_y s^2_z - s^2_y c^2_z \right )
- c_y s_z \left ( s^2_x s^2_y e^{+i\delta} + c^2_x c^2_y e^{-i\delta} \right )}
{s_x c_x s_y c^2_y + 
\left ( s^2_x - c^2_x \right ) c^3_y s_z e^{i\delta} + 
s_x c_x s_y s^2_z \left ( 1 + c^2_y \right ) e^{2i\delta}} ~ e^{2i\delta} \cr
\frac{\lambda_2}{\lambda_3} = 
\frac{s_x c_x s_y \left (2 c^2_y s^2_z - s^2_y c^2_z \right )
+ c_y s_z \left ( c^2_x s^2_y e^{+i\delta} + s^2_x c^2_y e^{-i\delta} \right )}
{s_x c_x s_y c^2_y + 
\left ( s^2_x - c^2_x \right ) c^3_y s_z e^{i\delta} + 
s_x c_x s_y s^2_z \left ( 1 + c^2_y \right ) e^{2i\delta}} ~ e^{2i\delta} }$
}
\\ \\
$\rm B_2: 
\left ( \matrix{
\times & {\bf 0} & \times \cr
{\bf 0} & \times & \times \cr
\times & \times & {\bf 0} \cr} \right )$
&& 
{\large
$\matrix{
\frac{\lambda_1}{\lambda_3} =  
\frac{s_x c_x c_y \left (2 s^2_y s^2_z - c^2_y c^2_z \right )
+ s_y s_z \left ( s^2_x c^2_y e^{+i\delta} + c^2_x s^2_y e^{-i\delta} \right )}
{s_x c_x s^2_y c_y - 
\left ( s^2_x - c^2_x \right ) s^3_y s_z e^{i\delta} + 
s_x c_x c_y s^2_z \left ( 1 + s^2_y \right ) e^{2i\delta}} ~ e^{2i\delta} \cr
\frac{\lambda_2}{\lambda_3} = 
\frac{s_x c_x c_y \left (2 s^2_y s^2_z - c^2_y c^2_z \right )
- s_y s_z \left ( c^2_x c^2_y e^{+i\delta} + s^2_x s^2_y e^{-i\delta} \right )}
{s_x c_x s^2_y c_y - 
\left ( s^2_x - c^2_x \right ) s^3_y s_z e^{i\delta} + 
s_x c_x c_y s^2_z \left ( 1 + s^2_y \right ) e^{2i\delta}} ~ e^{2i\delta} }$
}
\\ \\
$\rm B_3: 
\left ( \matrix{
\times & {\bf 0} & \times \cr
{\bf 0} & {\bf 0} & \times \cr
\times & \times & \times \cr} \right )$
&& 
{\large
$\matrix{
\frac{\lambda_1}{\lambda_3} =  
- \frac{s_y}{c_y} \cdot \frac{s_x s_y - c_x c_y s_z e^{-i\delta}}
{s_x c_y + c_x s_y s_z e^{+i\delta}} ~ e^{2i\delta} \cr
\frac{\lambda_2}{\lambda_3} =  
- \frac{s_y}{c_y} \cdot \frac{c_x s_y + s_x c_y s_z e^{-i\delta}}
{c_x c_y - s_x s_y s_z e^{+i\delta}} ~ e^{2i\delta} }$
}
\\ \\
$\rm B_4: 
\left ( \matrix{
\times & \times & {\bf 0} \cr
\times & \times & \times \cr
{\bf 0} & \times & {\bf 0} \cr} \right )$
&&
{\large 
$\matrix{
\frac{\lambda_1}{\lambda_3} =  
- \frac{c_y}{s_y} \cdot \frac{s_x c_y + c_x s_y s_z e^{-i\delta}}
{s_x s_y - c_x c_y s_z e^{+i\delta}} ~ e^{2i\delta} \cr
\frac{\lambda_2}{\lambda_3} =  
- \frac{c_y}{s_y} \cdot \frac{c_x c_y - s_x s_y s_z e^{-i\delta}}
{c_x s_y + s_x c_y s_z e^{+i\delta}} ~ e^{2i\delta} }$
}
\\ \\
$\rm C: ~
\left ( \matrix{
\times & \times & \times \cr
\times & {\bf 0} & \times \cr
\times & \times & {\bf 0} \cr} \right )$
&& 
{\large
$\matrix{
\frac{\lambda_1}{\lambda_3} = 
- \frac{c_x c^2_z}{s_z} \cdot \frac{ c_x \left ( s^2_y - c^2_y \right ) 
+ 2 s_x s_y c_y s_z e^{i\delta}}
{2 s_x c_x s_y c_y - \left ( s^2_x - c^2_x \right )
\left ( s^2_y - c^2_y \right ) s_z e^{i\delta} + 2 s_x c_x s_y c_y s^2_z
e^{2i\delta}} ~ e^{i\delta} \cr
\frac{\lambda_2}{\lambda_3} = 
+ \frac{s_x c^2_z}{s_z} \cdot \frac{ s_x \left ( s^2_y - c^2_y \right ) 
- 2 c_x s_y c_y s_z e^{i\delta}}
{2 s_x c_x s_y c_y - \left ( s^2_x - c^2_x \right )
\left ( s^2_y - c^2_y \right ) s_z e^{i\delta} + 2 s_x c_x s_y c_y s^2_z
e^{2i\delta}} ~ e^{i\delta} }$
}
\\ \\ 
$\rm D_1:
\left ( \matrix{
\times & \times & \times \cr
\times & {\bf 0} & {\bf 0} \cr
\times & {\bf 0} & \times \cr} \right )$
&& 
{\large
$\matrix{
\frac{\lambda_1}{\lambda_3} = 
- \frac{c^2_z}{s_z} \cdot \frac{c_x s_y}{s_x c_y + 
c_x s_y s_z e^{i\delta}} ~ e^{i\delta} \cr
\frac{\lambda_2}{\lambda_3} = 
+ \frac{c^2_z}{s_z} \cdot \frac{s_x s_y}{c_x c_y -
s_x s_y s_z e^{i\delta}} ~ e^{i\delta} }$
}
\\ \\
$\rm D_2:
\left ( \matrix{
\times & \times & \times \cr
\times & \times & {\bf 0} \cr
\times & {\bf 0} & {\bf 0} \cr} \right )$
&& 
{\large
$\matrix{
\frac{\lambda_1}{\lambda_3} = 
+ \frac{c^2_z}{s_z} \cdot \frac{c_x c_y}{s_x s_y -
c_x c_y s_z e^{i\delta}} ~ e^{i\delta} \cr
\frac{\lambda_2}{\lambda_3} = 
- \frac{c^2_z}{s_z} \cdot \frac{s_x c_y}{c_x s_y +
s_x c_y s_z e^{i\delta}} ~ e^{i\delta} }$
}
\\ \\ \hline\hline
\end{tabular}
\end{center}
\end{table}

\begin{table}
\caption{Six patterns of the neutrino mass matrix $M$ with two independent
vanishing entries, which are {\it incompatible} with the LOW solution and other
empirical hypotheses. The analytical results for two ratios of three neutrino 
mass eigenvalues $\lambda_1/\lambda_3$ and $\lambda_2/\lambda_3$ are 
given in terms of four flavor mixing parameters $\theta_x$, $\theta_y$, 
$\theta_z$ and $\delta$.}
\begin{center}
\begin{tabular}{lcl} \hline\hline 
Pattern of $M$ &~~& Results of $\lambda_1/\lambda_3$ and 
$\lambda_2/\lambda_3$ \\ \hline \\
$\rm E_1: 
\left ( \matrix{
{\bf 0} & \times & \times \cr
\times  & {\bf 0} & \times \cr
\times & \times & \times \cr} \right )$
&& 
{\large
$\matrix{
\frac{\lambda_1}{\lambda_3} =  
- \frac{1}{c_y c^2_z} \cdot \frac{s^2_x s^2_y \left (c^2_z - s^2_z \right )
- c_x c_y s^2_z \left (c_x c_y - 2 s_x s_y s_z e^{i\delta} \right ) e^{-2i\delta}}
{\left (s^2_x - c^2_x \right ) c_y + 2 s_x c_x s_y s_z e^{i\delta}} 
~ e^{2i\delta} \cr
\frac{\lambda_2}{\lambda_3} =  
+ \frac{1}{c_y c^2_z} \cdot \frac{c^2_x s^2_y \left (c^2_z - s^2_z \right )
- s_x c_y s^2_z \left (s_x c_y + 2 c_x s_y s_z e^{i\delta} \right ) e^{-2i\delta}}
{\left (s^2_x - c^2_x \right ) c_y + 2 s_x c_x s_y s_z e^{i\delta}}
~ e^{2i\delta} }$
}
\\ \\
$\rm E_2: 
\left ( \matrix{
{\bf 0} & \times & \times \cr
\times & \times & \times \cr
\times & \times & {\bf 0} \cr} \right )$
&& 
{\large
$\matrix{
\frac{\lambda_1}{\lambda_3} =  
- \frac{1}{s_y c^2_z} \cdot \frac{s^2_x c^2_y \left (c^2_z - s^2_z \right )
- c_x s_y s^2_z \left (c_x s_y + 2 s_x c_y s_z e^{i\delta} \right ) e^{-2i\delta}}
{\left (s^2_x - c^2_x \right ) s_y - 2 s_x c_x c_y s_z e^{i\delta}} 
~ e^{2i\delta} \cr
\frac{\lambda_2}{\lambda_3} =  
+ \frac{1}{s_y c^2_z} \cdot \frac{c^2_x c^2_y \left (c^2_z - s^2_z \right )
- s_x s_y s^2_z \left (s_x s_y - 2 c_x c_y s_z e^{i\delta} \right ) e^{-2i\delta}}
{\left (s^2_x - c^2_x \right ) s_y - 2 s_x c_x c_y s_z e^{i\delta}}
~ e^{2i\delta} }$
}
\\ \\
$\rm E_3: 
\left ( \matrix{
{\bf 0} & \times & \times \cr
\times & \times & {\bf 0} \cr
\times & {\bf 0} & \times \cr} \right )$
&& 
{\large
$\matrix{
\frac{\lambda_1}{\lambda_3} =  
- \frac{1}{c^2_z} \cdot \frac{s^2_x s_y c_y \left (c^2_z - s^2_z \right )
+ c_x s^2_z \left [c_x s_y c_y - s_x \left (s^2_y - c^2_y \right ) s_z e^{i\delta} 
\right ] e^{-2i\delta}}
{\left (c^2_x - s^2_x \right ) s_y c_y + s_x c_x \left (c^2_y - s^2_y \right )
s_z e^{i\delta}} ~ e^{2i\delta} \cr
\frac{\lambda_2}{\lambda_3} = 
+ \frac{1}{c^2_z} \cdot \frac{c^2_x s_y c_y \left (c^2_z - s^2_z \right )
+ s_x s^2_z \left [s_x s_y c_y + c_x \left (s^2_y - c^2_y \right ) s_z e^{i\delta} 
\right ] e^{-2i\delta}}
{\left (c^2_x - s^2_x \right ) s_y c_y + s_x c_x \left (c^2_y - s^2_y \right )
s_z e^{i\delta}} ~ e^{2i\delta} }$
}
\\ \\
$\rm F_1: 
\left ( \matrix{
\times & {\bf 0} & {\bf 0} \cr
{\bf 0} & \times & \times \cr
{\bf 0} & \times & \times \cr} \right )$
&& 
{\large
$\matrix{
\frac{\lambda_1}{\lambda_3} = 1 \cr
\frac{\lambda_2}{\lambda_3} = 1 }$
}
\\ \\
$\rm F_2: 
\left ( \matrix{
\times & {\bf 0} & \times \cr
{\bf 0} & \times & {\bf 0} \cr
\times & {\bf 0} & \times \cr} \right )$
&& 
{\large
$\matrix{
\frac{\lambda_1}{\lambda_3} =  
\frac{s_x c_y + c_x s_y s_z e^{-i\delta}}
{s_x c_y + c_x s_y s_z e^{+i\delta}} ~ e^{2i\delta} \cr
\frac{\lambda_2}{\lambda_3} =  
\frac{c_x c_y - s_x s_y s_z e^{-i\delta}}
{c_x c_y - s_x s_y s_z e^{+i\delta}} ~ e^{2i\delta} }$
}
\\ \\
$\rm F_3: 
\left ( \matrix{
\times & \times & {\bf 0} \cr
\times & \times & {\bf 0} \cr
{\bf 0} & {\bf 0} & \times \cr} \right )$
&&
{\large 
$\matrix{
\frac{\lambda_1}{\lambda_3} =  
\frac{s_x s_y - c_x c_y s_z e^{-i\delta}}
{s_x s_y - c_x c_y s_z e^{+i\delta}} ~ e^{2i\delta} \cr
\frac{\lambda_2}{\lambda_3} =  
\frac{c_x s_y + s_x c_y s_z e^{-i\delta}}
{c_x s_y + s_x c_y s_z e^{+i\delta}} ~ e^{2i\delta} }$
}
\\ \\ \hline\hline
\end{tabular}
\end{center}
\end{table}

\newpage

\begin{figure}[t]
\vspace{-1cm}
\epsfig{file=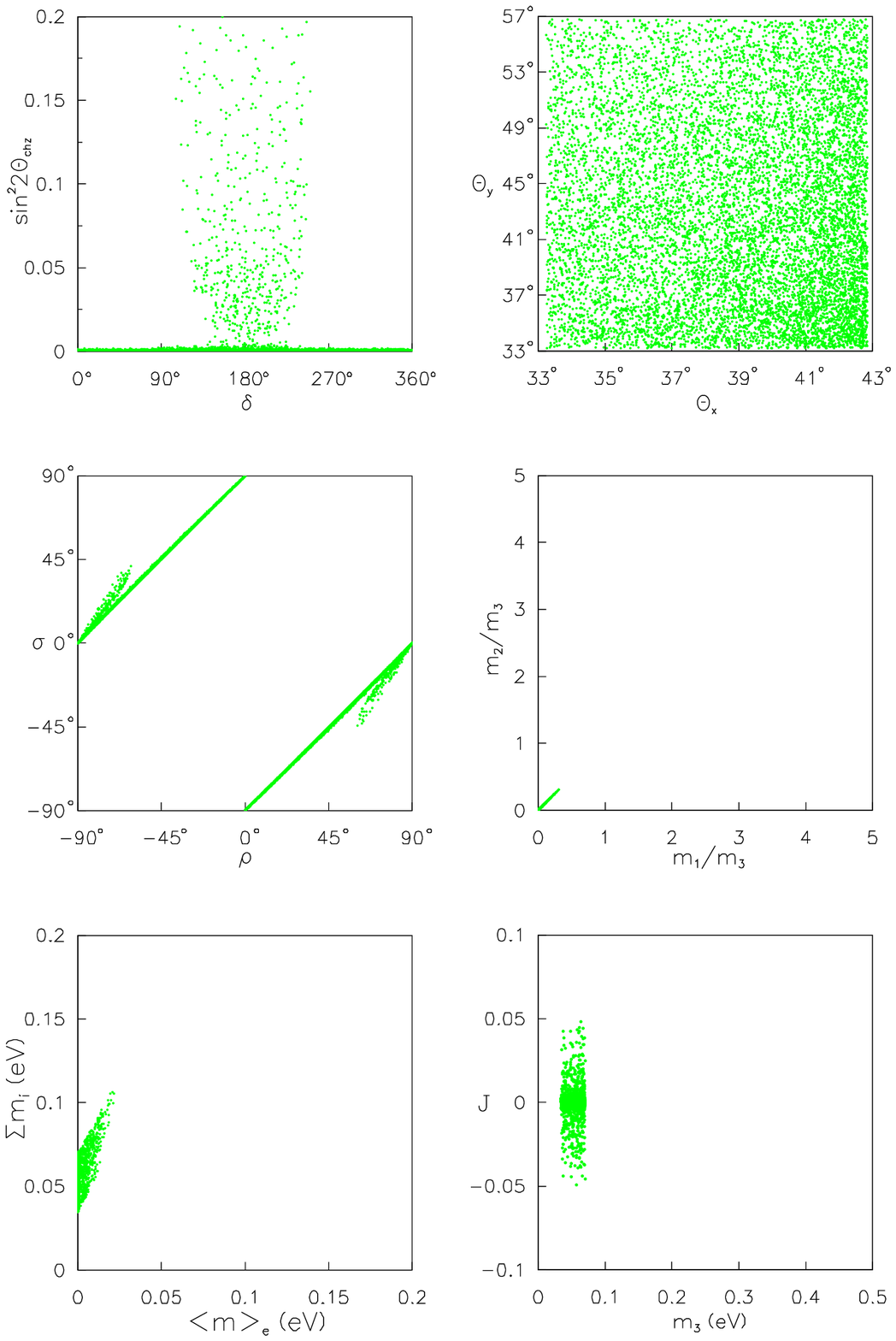,bbllx=2cm,bblly=2cm,bburx=17.5cm,bbury=28cm,%
width=16cm,height=26cm,angle=0,clip=0} 
\vspace{-4.5cm} 
\caption{Pattern $\rm A_1$ of the neutrino mass matrix $M$: allowed regions 
of $\sin^2 2\theta_{\rm chz}$ versus $\delta$, $\theta_y$ versus $\theta_x$, 
$\sigma$ versus $\rho$, $m_2/m_3$ versus $m_1/m_3$,
$\sum m_i$ versus $\langle m \rangle_e$, and $J$ versus $m_3$.}
\end{figure}

\begin{figure}[t]
\vspace{-1cm}
\epsfig{file=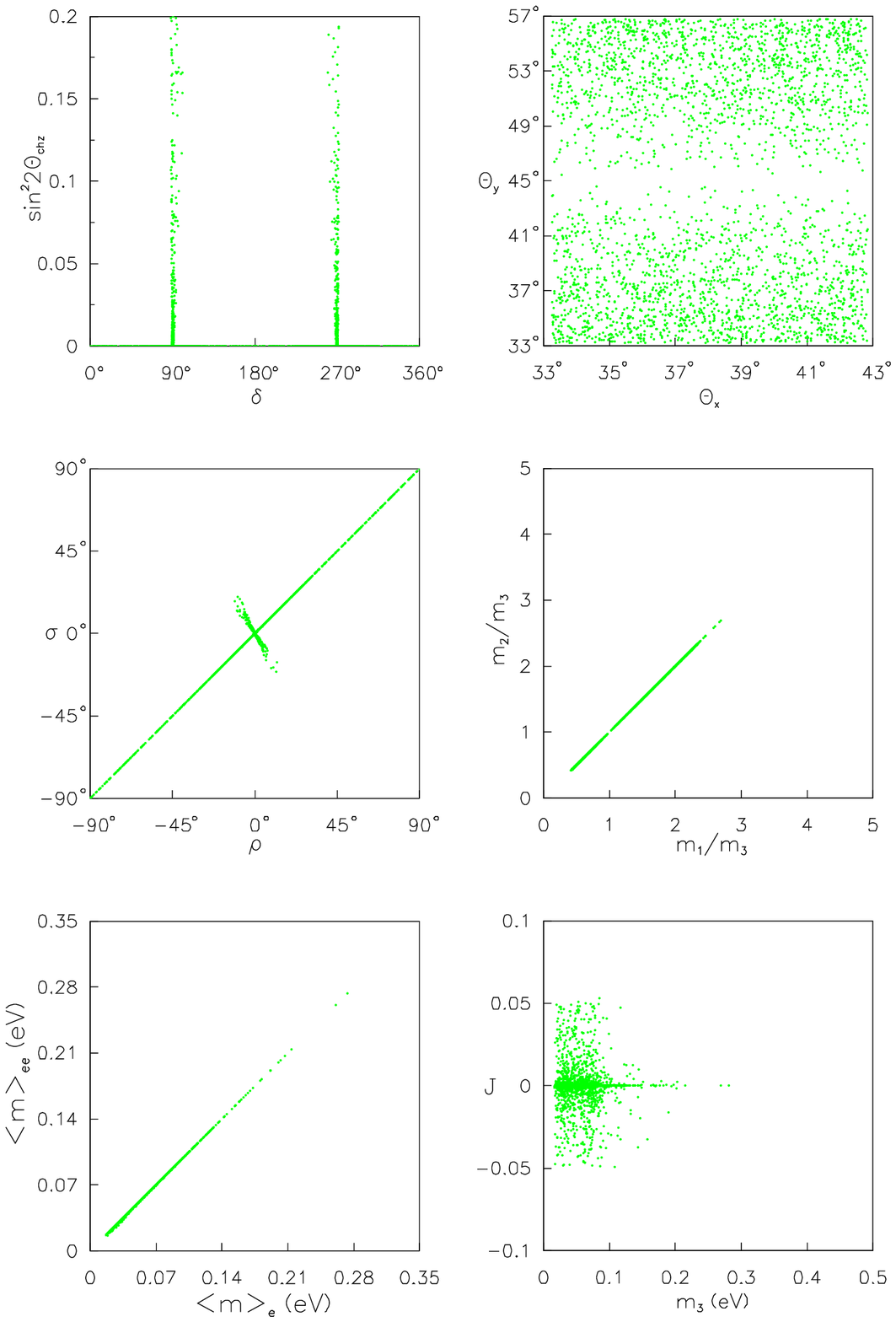,bbllx=2cm,bblly=2cm,bburx=17.5cm,bbury=28cm,%
width=16cm,height=26cm,angle=0,clip=0} 
\vspace{-4.5cm} 
\caption{Pattern $\rm B_1$ of the neutrino mass matrix $M$: allowed regions 
of $\sin^2 2\theta_{\rm chz}$ versus $\delta$, $\theta_y$ versus $\theta_x$, 
$\sigma$ versus $\rho$, $m_2/m_3$ versus $m_1/m_3$,
$\langle m \rangle_{ee}$ versus $\langle m \rangle_e$, and $J$ 
versus $m_3$.}
\end{figure}

\begin{figure}[t]
\vspace{-1cm}
\epsfig{file=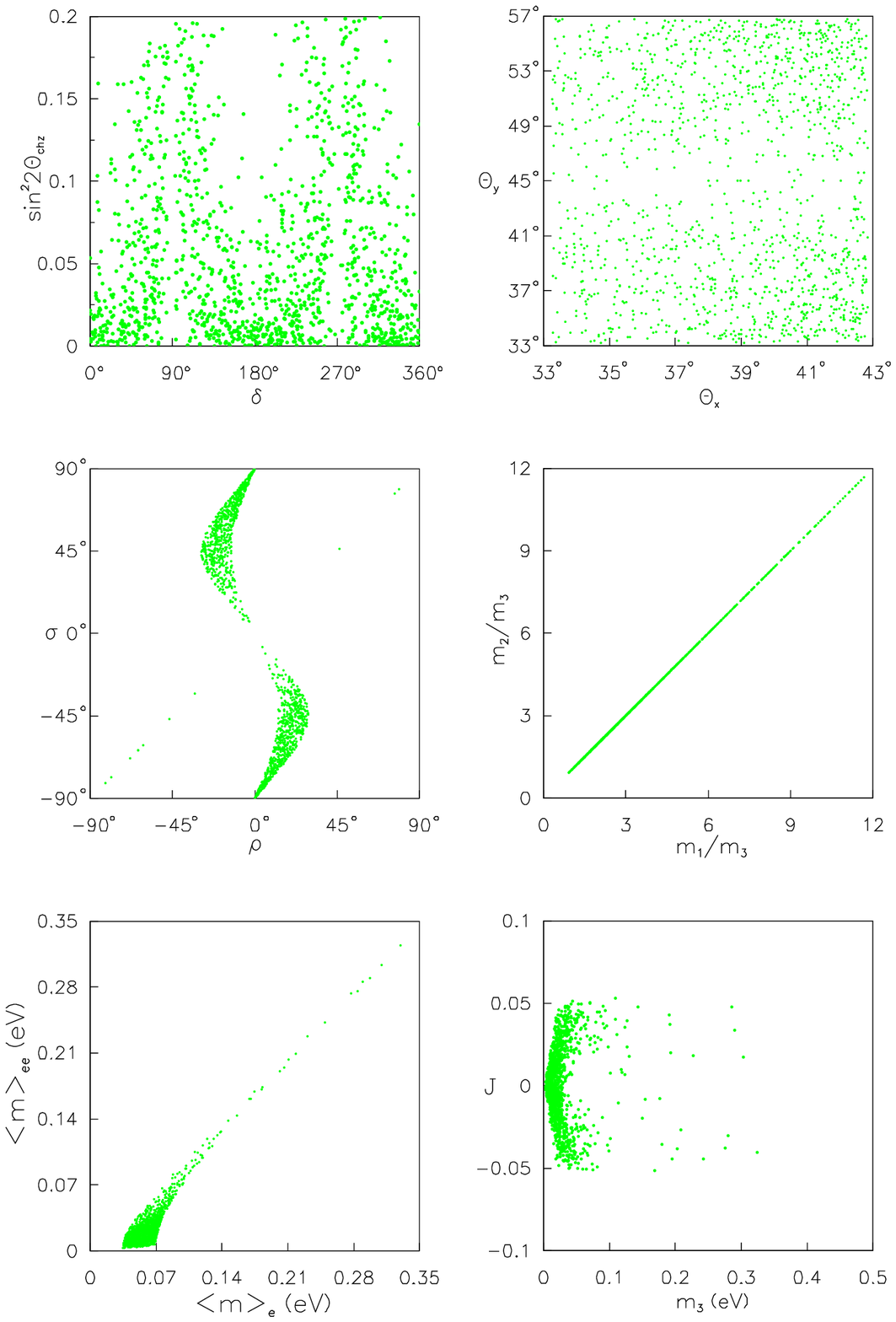,bbllx=2cm,bblly=2cm,bburx=17.5cm,bbury=28cm,%
width=16cm,height=26cm,angle=0,clip=0} 
\vspace{-4.5cm} 
\caption{Pattern $\rm C$ of the neutrino mass matrix $M$: allowed regions 
of $\sin^2 2\theta_{\rm chz}$ versus $\delta$, $\theta_y$ versus $\theta_x$, 
$\sigma$ versus $\rho$, $m_2/m_3$ versus $m_1/m_3$,
$\langle m \rangle_{ee}$ versus $\langle m \rangle_e$, and $J$ 
versus $m_3$.}
\end{figure}

\begin{figure}[t]
\vspace{-1cm}
\epsfig{file=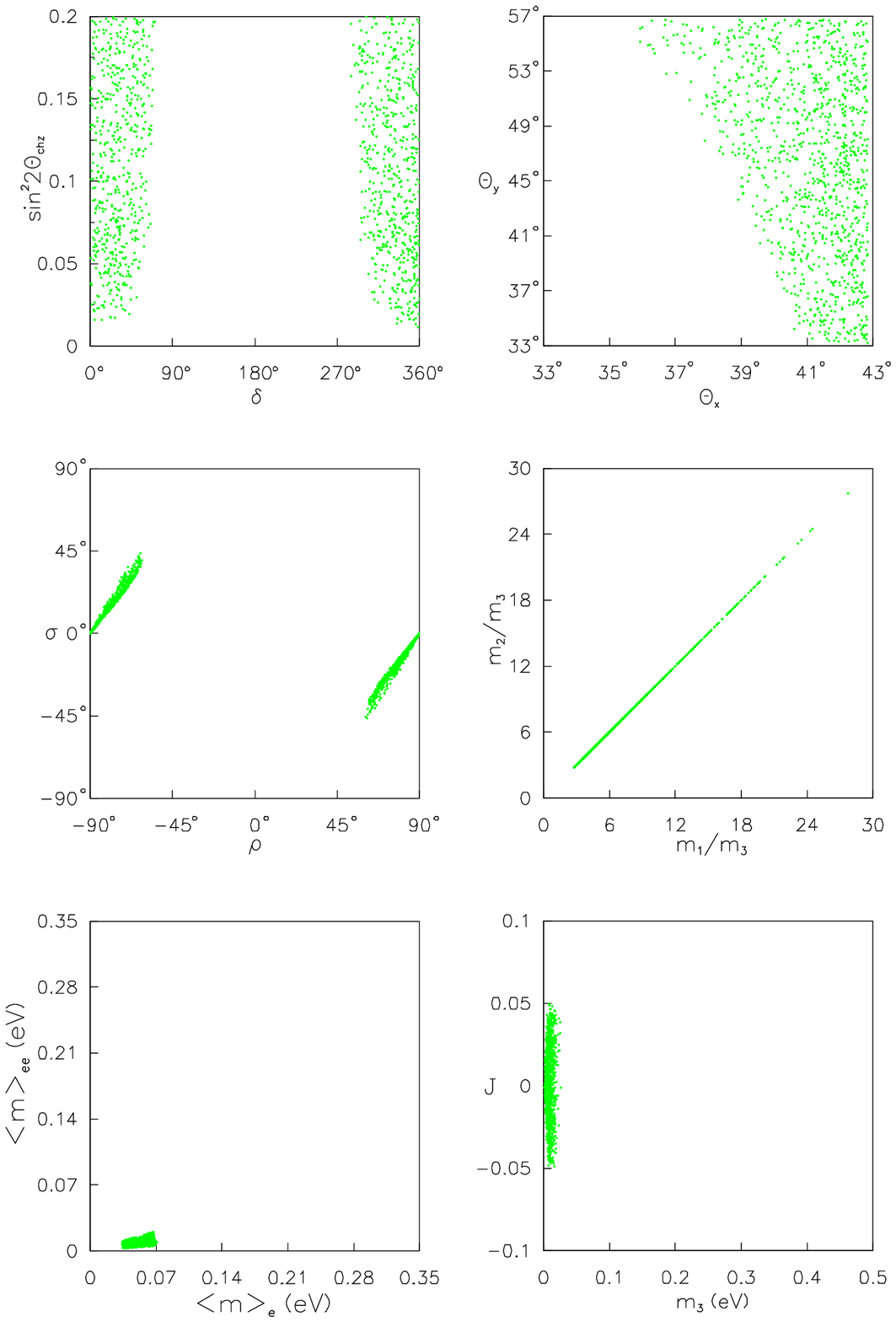,bbllx=2cm,bblly=2cm,bburx=17.5cm,bbury=28cm,%
width=16cm,height=26cm,angle=0,clip=0} 
\vspace{-4.5cm} 
\caption{Pattern $\rm D_1$ of the neutrino mass matrix $M$: allowed regions 
of $\sin^2 2\theta_{\rm chz}$ versus $\delta$, $\theta_y$ versus $\theta_x$, 
$\sigma$ versus $\rho$, $m_2/m_3$ versus $m_1/m_3$,
$\langle m \rangle_{ee}$ versus $\langle m \rangle_e$, and $J$ 
versus $m_3$.}
\end{figure}

\end{document}